\documentclass{article}
\usepackage[T1]{fontenc} 
\usepackage[utf8]{inputenc} 
\usepackage{ismir,amsmath,cite,url}
\usepackage{graphicx}
\usepackage{xcolor}
\usepackage{booktabs}
\usepackage{amsfonts}
\usepackage{enumitem}
\usepackage{tablefootnote}
\usepackage{lineno}
\usepackage{stfloats}
\usepackage{musicography}
\usepackage{makecell}
\usepackage[bookmarks=false, colorlinks=true, linkcolor=black, citecolor=black, urlcolor=black]{hyperref}
\usepackage{array,etoolbox}
\preto\tabular{\setcounter{magicrownumbers}{0}}
\newcounter{magicrownumbers}
\def\rownumber{}

\newcounter{mycounter}
\newcommand{\rowlabel}[1]{\refstepcounter{mycounter} \label{#1}}
\title{End-to-End Piano Performance-MIDI to Score Conversion with Transformers}

\twoauthors
  {Tim Beyer} {Technical University of Munich \\ {\tt tim.beyer@tum.de}}
  {Angela Dai} {Technical University of Munich \\ {\tt angela.dai@tum.de}}

\def\authorname{T. Beyer and A. Dai}

\begin{document}

\maketitle
\begin{abstract}
\vspace{-0.4em}
The automated creation of accurate musical notation from an expressive human performance is a fundamental task in computational musicology.
To this end, we present an end-to-end deep learning approach that constructs detailed musical scores directly from real-world piano performance-MIDI files. 
We introduce a modern transformer-based architecture with a novel tokenized representation for symbolic music data.
Framing the task as sequence-to-sequence translation rather than note-wise classification reduces alignment requirements and annotation costs, while allowing the prediction of more concise and accurate notation.
To serialize symbolic music data, we design a custom tokenization stage based on compound tokens that carefully quantizes continuous values. 
This technique preserves more score information while reducing sequence lengths by $3.5\times$ compared to prior approaches.
Using the transformer backbone, our method demonstrates better understanding of note values, rhythmic structure, and details such as staff assignment.
When evaluated end-to-end using transcription metrics such as MUSTER, we achieve significant improvements over previous deep learning approaches and complex HMM-based state-of-the-art pipelines.
Our method is also the first to directly predict notational details like trill marks or stem direction from performance data.
Code and models are available on \href{https://github.com/TimFelixBeyer/MIDI2ScoreTransformer}{\color{cyan}GitHub}.

\end{abstract}
\vspace{-0.6em}
\section{Introduction}\label{sec:introduction}
Creating structured musical scores from human performance recordings is a challenging task with a significant number of downstream applications in areas such as alignment \cite{orio2001alignment}, score-following, education, and archiving.

Human performances are typically represented as performance-MIDI (P-MIDI) files, as they can be easily recorded from MIDI instruments or generated from audio by automated transcription systems \cite{hawthorne2017onsets,benetos2018automatic}.
In contrast, high-quality scores in standard sheet music formats such as MusicXML \cite{good2001musicxml} are much less commonly available and generally require creation by human experts.

Performance-MIDI-to-Score conversion (PM2S) is complex, encompassing several lower-level tasks like note value prediction, tempo regression, rhythm quantization, voice assignment, and typesetting details, including ornaments and note stems. 
As a result, PM2S and its sub-tasks have remained an active research topic and popular application of computational methods for over 30 years \cite{desain1989quantization}.

While early approaches relied on classical modeling and hand-crafted processing \cite{raphael2001automated}, research gradually shifted towards statistical methods based on Hidden Markov Models (HMMs) augmented with heuristics.

Cogliati et al. \cite{cogliati2016transcribing} run an HMM-based meter estimation \cite{temperley2009unified} with beat-snapping heuristics to quantize note timings, before outputting LilyPond \cite{nienhuys2003lilypond} notation.
HMM variants were also used for estimating staff placement \cite{nakamura2014merged} and rhythm quantization \cite{nakamura2017rhythm,nakamura2018towards}, where the outputs of multiple models are merged into a final prediction with improved accuracy.
To complement onset timing quantization, a method for note value recognition based on Markov random fields was introduced by Nakamura et al. \cite{nakamura2017note}.
Building upon these advances, Shibata et al. \cite{shibata2021non} combined prior systems with hand-crafted non-local statistics to improve estimates of global attributes such as piece tempo and time signatures. 
While these approaches yield state-of-the-art performance, they are composed of a complex web of interdependent components, rely on human-designed priors, and are not trained end-to-end. 
Thus, more recent work has attempted to tackle PM2S using deep learning.

So far, the scarcity of high-quality labeled data has limited its use to scenarios with cheaper labels, leading to a focus on synthetic data \cite{carvalho2017towards,roman2020data}, sub-tasks like pitch-spelling \cite{foscarin2021pkspell}, note value quantization and voicing \cite{hiramatsu2021joint}, or beat tracking \cite{liu2022performance,cheng2023transformer}.
Beat tracking, in particular, has seen significant progress by combining CRNNs with beat in-filling via dynamic programming \cite{liu2022performance}. 
The CRNN also predicts other score attributes such as key and time signatures.
Unfortunately, beat-tracking makes overly restrictive assumptions about the regularity of the underlying performance data and struggles with real-world human recordings.

Another challenge is the representation of symbolic music data for machine learning models.
Monophonic sequences are often captured as Lilypond \cite{nienhuys2003lilypond,carvalho2017towards}, ABC \cite{walshaw2011abc}, Humdrum-derived \cite{huron2002music,roman2020data}, or custom CTC-friendly \cite{roman2020data} character sequences. 
For polyphonic data, piano rolls \cite{kelz2016potential,hawthorne2017onsets}, MIDI-derived tokens \cite{huang2020pop,hawthorne2021sequence,gardner2021mt3}, and custom MusicXML tokens \cite{suzuki2021score} are the most common representations.

To create more compact encodings, Zeng et al. \cite{zeng2021musicbert} and Dong et al. \cite{dong2023multitrack} use compound tokens and represent MIDI attributes in separate streams, shortening sequence lengths.

Our proposed approach continues the progression of PM2S systems towards end-to-end learned approaches and overcomes several limitations of prior systems based on deep learning, making the following key contributions:
\vspace{-0.1em}
\begin{itemize}[leftmargin=0.5cm,noitemsep]
    \item We cast PM2S as an end-to-end sequence-to-sequence translation task, developing a transformer to 
 enable accurate prediction of global attributes (e.g., meter) that require understanding of long-term dependencies. 
    \item Relaxed annotation requirements compared to prior deep learning methods, using only beat-level alignment for training. We can additionally leverage unmatched MusicXML data without corresponding P-MIDI. 
    \item We introduce a compact and extensible tokenization scheme for P-MIDI and MusicXML data, allowing the backbone model to directly translate tokenized P-MIDI into MusicXML tokens and enabling the generation of detailed score features such as ornaments.
    \item We demonstrate superior performance on quantitative error metrics like MUSTER, with our approach surpassing prior deep learning models and the highly optimized, complex state-of-the-art. 
\end{itemize}
\vspace{-1.1em}
\section{Methodology}
\subsection{Task definition}
Our end-to-end PM2S system directly converts an unstructured P-MIDI file into a highly readable MusicXML score. 
P-MIDI files only contain information about note timing (onsets, offsets), pitch, and velocity. 
The input to a PM2S system is thus defined by the following sequence:
\vspace{-0.3em}
\begin{equation}
    \mathbf{X} = \{\left(p_i, o_i, d_i, v_i\right)\}_{i=1}^{N_{\text{perf}}},
    \label{eq:input}
\end{equation}
with MIDI pitch $p_i$, onset $o_i$ and duration $d_i$ in seconds, and velocity $v_i$ for each of the $N_\text{perf}$ performance notes.

In contrast to existing methods, which often cast PM2S as a note-wise classification task \cite{liu2022performance,hiramatsu2021joint}, we do not assume a one-to-one correspondence between notes in the performance and the score.
This is crucial in scenarios with trills or misplayed notes, where one-to-one matchings are impossible.
Consequently, we predict a new output note sequence from scratch that includes a full set of MusicXML attributes for each note in the score:
\vspace{-0.4em}
\begin{equation}
\vspace{-0.3em}
    \mathbf{Y_q} = \{\left(p_j, mo_j, md_j, ml_j \right)\}_{j=1}^{N_{\text{score}}}
    \label{eq:y_q}
\end{equation}
\begin{equation}
\vspace{-0.15em}
    \mathbf{Y_v} = \{\left(h_j, vo_j \right)\}_{j=1}^{N_{\text{score}}}
\end{equation}
\begin{equation}
    \mathbf{Y_o} = \{\left(t_j, s_j, sd_j, g_j, a_j \right)\}_{j=1}^{N_{\text{score}}}
    \label{eq:y_o}
\end{equation}
\begin{equation}
\vspace{-0.15em}
    \mathbf{Y} = (\mathbf{Y_q}, \mathbf{Y_v}, \mathbf{Y_o}),
   \label{eq:y}
\end{equation}
where $\mathbf{Y_q}$ comprises attributes related to pitch $p_j$ and quantized timings for the musical onset time $mo_j$, musical duration $md_j$, and measure length $ml_j$.
$\mathbf{Y_v}$ collects vertical positioning information, such as staff placement/hand $h_j$, and MusicXML voice number $vo_j$.
Finally, $\mathbf{Y_o}$ covers performance annotations, ornamentation, and typesetting details like trill $t_j$, staccato $s_j$, stem direction $sd_j$, grace note $g_j$, and accidentals $a_j$.
Predicting these additional attributes enables creating more concise and accurate notation.
$\mathbf{X}$ and $\mathbf{Y}$ are sorted by ascending onset/offset, pitch, and duration, yielding a unique serialized representation even for complex polyphony.

\subsection{Tokenization scheme}
\vspace{-0.3em}
To efficiently represent input $\mathbf{X}$ and output $\mathbf{Y}$, we introduce a systematic tokenization for P-MIDI files and MusicXML scores.
The key objective of a tokenization algorithm is to retain as much information from the original sequence as possible within a compact sequence length and vocabulary size.
Thus, we adopt a parallel token stream paradigm \cite{zeng2021musicbert}; a separate token stream is constructed for each of the four input attributes given in Eq.~\eqref{eq:input} and the eleven output attributes in Eq.~\eqref{eq:y}.
As a result, each note occupies only one timeslot.
The final vocabulary sizes and parameter ranges are shown in \tabref{tab:parameter-specifications}. 

For P-MIDI, we adopt a strategy similar to \cite{liu2022performance} and use 128 pitch tokens, 8 quantized velocity tokens, and quantize delta onsets and durations into 200 buckets.
To achieve high resolution for small values while covering times up to 8 seconds without clipping, we apply a $log$-transform before bucketing onsets and durations, implementing a continuous version of multi-resolution quantization \cite{zeng2021musicbert}.

We pay particular attention to the MusicXML tokenization. 
While binary and categorical attributes, such as stem direction and staff assignment, are easily tokenized, continuous values like onsets and durations demand more care.
The encoding of musical timing and positioning significantly impacts score quality. 
To find a good trade-off between vocabulary size and the ability to correctly represent common note durations and onsets, we conducted a search over bucket sizes.
Consequently, we opt to quantize musical time into $\frac{1}{24}$th fractions, diverging from previous approaches, which rely on powers of 2 or smaller denominators \cite{hiramatsu2021joint,zeng2021musicbert,lv2023getmusic}.
Our parameterization accurately represents 98.6\% of notes in the ASAP dataset with 97 tokens, compared to just 85.4\% using powers of 2 up to 256. 

Encoding absolute musical onsets into tokens poses another challenge.
Direct quantization of absolute positions is infeasible due to the large range of positions required, while delta encoding similar to MIDI quickly leads to drift issues and misalignment with \hspace{-0.2cm}
\begin{figure*}[!ht]
    \centering
        \includegraphics[alt={A diagram of our model architecture, showing how encoder and decoder stacks predict the next tokens.},width=\textwidth]{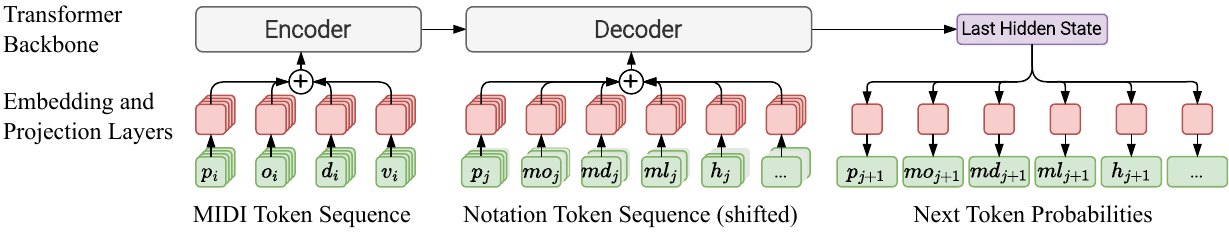}
        \vspace{-2.5em}
    \caption{Model architecture overview. We use a standard Roformer encoder-decoder model \cite{su2021roformer} with custom token embedding and projection layers. Each token stream is embedded separately, then a constant-size shared embedding is created via summation. The backbone model architecture remains unchanged compared to models applied to NLP or other sequence-to-sequence learning tasks. In this illustration, depth symbolizes the time direction.
    }
    \vspace{-1.2em}
    \label{fig:model-overview}
\end{figure*}
\begin{table}[!h]
\vspace{-0.7em}
\setlength{\tabcolsep}{3pt}
\centering
{\renewcommand{\arraystretch}{0.94}
\begin{tabular}{lrc}
\toprule
\textbf{Parameter} & $\boldsymbol{N_{vocab}}$ & \textbf{Range/Values} \vspace{-0.02cm}\\
\midrule
\multicolumn{3}{c}{\textit{Input Parameters}}\vspace{-0.05cm}\\
Pitch ($p_i$) & 128 &  \([0, 127]\) \\
Onset ($o_i$) & 200 &  \([0, 8]\) \\
Duration ($d_i$) & 200 & \([0, 8]\) \\
Velocity ($v_i$) & 8 &  \([0, 127]\)\vspace{-0.05cm}\\
\midrule
\multicolumn{3}{c}{\textit{Output Parameters}}\vspace{-0.05cm}\\
Pitch ($p_j$) & 128 & \([0, 127]\)\\
Musical Onset ($mo_j$) & 145 &  \([0, 6]\)\\
Musical Duration ($md_j$) & 97 &  \([0, 4]\)\\
Measure Length ($ml_j$) & 146 & \([0, 6] \cup \{\text{false}\}\)\\
\begin{tabular}{@{}l@{}}
     Hand/Staff ($h_j$), Trill ($t_j$), \\ Grace ($g_j$), Staccato ($s_j$)
\end{tabular} \hspace{-3em} & 2 each & boolean \\
Voice ($vo_j$) & 8 & \([1, 8]\)\\
Stem ($sd_j$) & 3 & \{up, down, none\}\\
Accidental ($a_j$) & 6 & \{\musDoubleFlat, \fl, \na, \sh, \musDoubleSharp, none\}\\
\bottomrule
\end{tabular}
}
\vspace{-0.45em}
\caption{Parameter specifications for input/output representations. The rightmost column details the range or set of representable values for each attribute. Continuous values outside the range are clipped before tokenization.}
\vspace{-0.2em}
\label{tab:parameter-specifications}
\setlength{\tabcolsep}{6pt}
\end{table}
measure boundaries and the musical grid. 
Thus, we adopt a hybrid approach representing absolute positions using two tokens; $mo_j$ encodes the note's position relative to the start of the current measure, and $ml_j$ stores the preceding measure's length for the first note in each measure or is set to \verb|false| otherwise.
Combined, $mo_j$ and $ml_j$ enable the reconstruction of absolute musical times, including correct bar lines and most time signatures.
During score creation, bar lines are used to split and tie notes crossing measure boundaries, recovering most ties, and rests are added to fill any gaps in voices.

Since input and output streams are not necessarily the same length, we also insert space tokens ($sp_j$) where required for alignment (see also Section \ref{subsubsec:training-batch-construction}).
The space tokens differ from typical end-of-sequence padding or masking tokens in Transformers, as they are predicted during inference and attention to these positions is not masked.
\vspace{-.6em}
\subsection{Model architecture}
\vspace{-.2em}
By adopting a unified autoregressive Transformer encoder-decoder model \cite{vaswani2017attention} to directly translate tokenized P-MIDI into MusicXML tokens (as depicted in \figref{fig:model-overview}), we diverge from existing deep learning models for PM2S, which used subtask-specific LSTMs \cite{hochreiter1997long} or CRNNs.
Our choice is driven by the transformer's ability to scale to large datasets and to handle long-range dependencies, which are crucial for predicting piece-wide attributes like meter.

To interface with parallel token streams, our model introduces custom embedding and projection modules.
First, each attribute-specific token stream is mapped into a constant-size 512-dimensional embedding space. 
The results are then summed and normalized using LayerNorm \cite{ba2016layer} to form a constant-size shared embedding, independent of the token stream count.

The backbone model itself follows the original architecture described by Vaswani et al. \cite{vaswani2017attention} and consists of symmetrically arranged encoder and decoder stacks. 
Each stack comprises four layers, eight attention heads, and a model dimension of 512.
To optimize performance, we adopt rotary positional encodings \cite{su2021roformer}, pre-norm \cite{brown2020language}, and SwiGLU activations \cite{shazeer2020glu} with an inner dimension of 3072 for the position-wise feed-forward network. 
At the end of the decoder, a set of linear layers projects the final hidden state into one output logit distribution per token stream.

\vspace{-0.6em}
\subsection{Training \& inference details}\label{sec:training-inference-details}
\vspace{-0.2em}
To optimize our model parameters, we break down the loss computation into two stages and first compute per-timestep losses $\mathcal{L}_j$, before summing along the sequence position.
At each timestep $j$, our model performs 12 separate classification tasks, one for every token stream in $\mathbf{Y}$ and one for the space token $sp_j$.

We compute the cross entropy (CE) loss for each output token stream $y$ and the space token stream.
For timesteps with spacing token $sp_j$, the loss is calculated only for the space token stream since the labels for all other tokens are undefined.
The full loss computation is thus:
\vspace{-0.3em}
\begin{align}
    \mathcal{L}_{y,j} &= \text{CE}\left(\hat{y}_j, y_j\right) \\
     \mathcal{L}_j &= 
    \begin{cases} 
        \text{CE}\left(\hat{sp}_j, 1\right)  &\text{for }sp_j = 1,\\
    \sum_{y\in{\mathbf{Y}}} \mathcal{L}_{y,j} + \text{CE}\left(\hat{sp}_j, 0\right)& \text{otherwise}.
    \end{cases}
    \\
    \mathcal{L} & = \sum_{j=1}^{N_\text{score}} \mathcal{L}_j.
\end{align}

\vspace{-0.3em}
We train our model for 40,000 steps using the AdamW optimizer \cite{loshchilov2017decoupled}.
The learning rate follows a cosine learning rate decay schedule with linear warmup over the first 4,000 steps to a maximum learning rate of 3e-4.
Gradients are clipped to a maximum value of 0.5. 
We use a batch size of 32 and the training sequence length is 512 timesteps. 

To parallelize training, transformers are often trained with teacher forcing. 
However, exposure bias \cite{bengio2015scheduled} can lead to lower-than-expected performance at inference time, especially in the low-data regime.
We find that heavy dropout \cite{srivastava2014dropout} during training to expose the model to only 25\% of the preceding output tokens addresses this problem.

During inference, we employ greedy top-1 decoding as it provides better performance than alternatives.
To handle songs with more than 512 notes, we partition the input into chunks of 512 notes each, ensuring a 64-note overlap between consecutive segments. 
Sufficient overlap eliminates abrupt changes at the segment boundaries and is essential for generating temporally coherent scores.
\vspace{-0.2em}
\section{Experiments}
\vspace{-0.2em}
\subsection{Data}
\subsubsection{Datasets} \label{subsub:datasets}
Unlike prior PM2S systems \cite{hiramatsu2021joint,shibata2021non,liu2022performance}, we do not use the MAPS \cite{emiya2010maps} dataset in our experiments, as performance data contained therein is not representative of real-world P-MIDI\footnote{The underlying data is score-derived and has been manually adjusted to represent aspects of performance and score at the same time. This leads to information leakage as highly regular onset/offset alignment remains in the `performance' data. Details of the laborious alignment process are available at \url{http://www.piano-midi.de/technic.htm}.}.
Furthermore, its musical scores are only available in the MIDI format, which lacks representational capacity compared to MusicXML. 
Thus, MAPS scores do not effectively capture many aspects of musical notation.

To overcome these limitations, we use the ASAP dataset \cite{foscarin2020asap} for training and evaluation.
It contains 1067 pieces of P-MIDI recorded from expert piano performances and corresponding high-fidelity MusicXML scores.
Performance and scores are aligned with beat-level annotations, which are significantly cheaper to obtain than note-level alignments.
We also observed that, on average, MusicXML scores contain 2.6\% fewer notes than associated P-MIDI.
These discrepancies are typically caused by misplayed notes and trills, again highlighting the importance of a flexible approach that is not reliant on one-to-one correspondences and can handle a wide variety of notation features.

After manually inspecting the dataset, we reject 100 instances due to poor alignment or data corruption, leaving 967 performances.
We perform only minimal preprocessing, focusing on removing non-sounding notes from the score.
This includes merging tied notes into a single, longer note and removing notes with the MusicXML \verb|print-object=no| attribute, as they would not be visible to a human performer.

To guarantee non-overlapping sets with robust evaluation across all composers in the dataset, dataset splits are created using the following procedure:

For each composer, we select one piece as a test piece and use all performances of this piece for the test set, yielding 59 instances.
90\% of all remaining pieces are used in the training set and 10\% in the validation set.
\tabref{tab:asap-statistics} shows the resulting full dataset split statistics.

To complement this labeled dataset, we also construct an unpaired dataset consisting of 58,646 public domain MusicXML files from \href{https://musescore.com}{Musescore}, without corresponding P-MIDI.
These scores are filtered for overlap with the labeled dataset to avoid data leakage.

\vspace{-0.5em}
\setlength{\tabcolsep}{3pt}
\begin{table}[!h]
\centering
\begin{tabular}{lrrrr}
\toprule
\textbf{Dataset} & \textbf{Train} & \textbf{Validation} & \textbf{Test} & \textbf{Total} \\
\midrule
\textbf{Performances} & 822 & 86 & 59 & 967 \\  
\textbf{Distinct pieces} & 176 & 16 & 14 & 206 \\
\textbf{P-MIDI Notes ($\boldsymbol{10^3}$)} & 2510 & 300 & 220 & 3030 \\
\textbf{Score Notes ($\boldsymbol{10^3}$)} & 2462 & 295 & 215 & 2972 \\
\bottomrule
\end{tabular}
\caption{Dataset statistics for ASAP \cite{foscarin2020asap} after excluding instances with mismatched annotations.}
\label{tab:asap-statistics}
\end{table}
\setlength{\tabcolsep}{6pt}
\vspace{-0.5em}

\subsubsection{Training batch construction}\label{subsubsec:training-batch-construction}
\vspace{-0.1em}
All training batches consist of 32 sequences of 512 notes each, equally split between labeled and unpaired datasets.
To sample instances from these heterogeneous datasets, we adopt two different procedures. 

\textbf{Labeled data.} We first use the beat-level correspondences to coarsely align input and output sequences by sorting notes into inter-beat intervals according to their onset time. 
Although this correspondence is exact for the MusicXML score data, human performances introduce variations to the P-MIDI data, causing some notes to not align perfectly with annotated beats. 
As a result, performance notes that occur shortly before the annotated beat time may musically belong into the next inter-beat interval and vice-versa.
To solve this issue, we follow a greedy optimization strategy that minimizes mismatched pitches between performance and score in each beat interval. 
If a performed note occurred within 50ms of a beat, and moving it to the previous/next inter-beat-interval reduces the number of mismatched pitches in both intervals, the move is performed. 
Where necessary for alignment, we add spacing tokens ($sp_j$) at the end of inter-beat intervals.
Given correct beat annotations, this procedure yields good alignment even in non-trivial situations like trills, where multiple MIDI notes correspond to just one MusicXML note. 

\textbf{Unpaired data.} In this case, only MusicXML data is available.
This could be used to simply pre-train the decoder stack in an autoregressive fashion; however, we found this procedure to be ineffective.
We thus aim to incorporate the encoder into the training process and construct a surrogate input token stream by reusing the output pitch tokens $p_j$ as input for the encoder model $p_i$ and mark the input sequence using conditioning tokens $c_i$. 
All other input tokens ($o_i$, $d_i$, and $v_i$) are masked.
As demonstrated in Section \ref{sec:ablation}, this significantly enhances the effectiveness of training on unpaired data.
Without input timing and velocity streams, the model has far less information to make predictions. 
To make the learning objective more feasible, we decrease the prior-token dropout probability to 50\% (compared to 75\% for paired data), improving training efficiency without compromising inference time behavior (see also Section \ref{sec:training-inference-details}). 
Similar to conditioning masks in diffusion models, we also feed a binary token ($c_i$) to the encoder which indicates that no real P-MIDI conditioning information from the labeled dataset is provided, resolving ambiguity about whether input tokens are masked/dropped out or simply not available.
The addition of this token improves the effectiveness of training on unlabeled data (see \tabref{tab:unified-ablation}).
When training on labeled data and during inference, its embedding is set to $\mathbf{0}$ and can thus be omitted.

\subsubsection{Data augmentation}
\vspace{-0.1em}
During training, four types of data augmentation are used to combat overfitting:
\begin{itemize}[noitemsep, leftmargin=0.5cm]
    \vspace{-0.8em}
    \item \textbf{Transposition:} Shift all pitches in the input and output up or down by up to 12 semitones; notes falling outside the MIDI pitch range are shifted inward by one octave. Accidentals are modified accordingly, following \cite{foscarin2021pkspell}.
    \item \textbf{Global tempo:} Change the timing data of the input MIDI notes by a factor of $\lambda \sim \mathcal{U}(0.8, 1.2)$.
    \item \textbf{Duration jitter:} To simulate human performance variations, performed note durations are additionally rescaled by a small amount of noise $\sim\mathcal{U}(0.95, 1.05)$.
    \item \textbf{Onset jitter:} All between-note intervals of the input MIDI are changed according to \\$\tilde{o}_{i+1} - \tilde{o}_i = (o_{i+1} - o_i) \cdot \mathcal{N}(1, {0.05}^2)$.
\vspace{-0.6em}
\end{itemize} 

\setlength{\tabcolsep}{2.7pt}

\begin{table*}[t]
\centering
\begin{tabular}{lrrrrrrrrrrrr}
\toprule
 & \multicolumn{6}{c}{\textbf{MUSTER} \cite{shibata2021non}} & \multicolumn{6}{c}{\textbf{ScoreSimilarity} \cite{cogliati2017metric,suzuki2021score}} \\
\cmidrule(lr){2-7} \cmidrule(lr){8-13}
\textbf{Method} & \( \boldsymbol{\mathcal{E}_\text{p}} \) & \( \boldsymbol{\mathcal{E}_\text{miss}} \)& \( \boldsymbol{\mathcal{E}_\text{extra}} \) & \( \boldsymbol{\mathcal{E}_\text{onset}} \) & \( \boldsymbol{\mathcal{E}_\text{offset}} \) & \( \boldsymbol{\mathcal{E}_\text{avg}} \) & \( \boldsymbol{\mathcal{E}_\text{miss}} \) & \( \boldsymbol{\mathcal{E}_\text{extra}} \) & \( \boldsymbol{\mathcal{E}_\text{dur.}} \) & \( \boldsymbol{\mathcal{E}_\text{staff}} \) & \( \boldsymbol{\mathcal{E}_\text{stem}} \) &  \( \boldsymbol{\mathcal{E}_\text{spell.}} \) \\
\midrule
Neural Beat Tracking (improved) \cite{liu2022performance} & \textbf{2.02} & 6.81 & 9.01 & 68.28 & 54.11 & 28.04 & 17.10 & 17.67 & 66.98 & 6.86 & - & 9.71 \\
MuseScore \cite{MuseScore2023} & 2.41 & 7.35 & 9.64 & 47.90 & 49.44 & 23.35 & 16.17 & 16.74 & 55.23 & 21.87 & 29.87 & 9.69\\
Finale \cite{FinaleV27} & 2.47 & 10.10 & 13.46 & 31.85 & 45.34 & 20.64 & 14.72 & 16.43 & 53.35 & 21.79 & 26.74 & 15.34 \\
HMMs + Heuristics (J-Pop) \cite{shibata2021non}$^\dagger$ & 2.09 & \textbf{6.38} & 8.67 & 25.02 & 29.21 & 14.27 & 10.80 & 11.39 & 71.38 & - & - & - \\ 
HMMs + Heuristics (classical) \cite{shibata2021non}$^\dagger$ & 2.11 & 6.47 & 8.75 & 22.58 & 29.84 & 13.95 & \textbf{10.74} & 11.28 & 64.73 & - & - & - \\
\midrule
\textbf{Ours} & 3.11 & 7.56 & \textbf{6.44} & \textbf{15.55} & \textbf{23.84} & \textbf{11.30} & 12.69 & \textbf{9.06} & \textbf{51.86} & \textbf{6.62} & \textbf{25.03} & \textbf{8.69} \\ 
\bottomrule
\end{tabular}
\vspace{-0.9em}
\caption{Comparative quantitative evaluation on the ASAP test set. All prior methods produce quantized MIDI and require MuseScore 4 to perform typesetting and conversion to MusicXML. $^\dagger$: the reported metrics are slightly optimistic as some pieces of the test set appeared in the training data for subcomponents of this method only.}
\vspace{-1.2em}
\label{tab:unified-comparative}
\end{table*}
\setlength{\tabcolsep}{6pt}
\subsection{Metrics}
To conduct fine-grained comparisons, we use both MUSTER \cite{nakamura2018towards,hiramatsu2021joint} and ScoreSimilarity \cite{cogliati2017metric,suzuki2021score} as evaluation metrics for PM2S performance\footnote{MV2H \cite{mcleod2019evaluating} was also considered, but its alignment procedure was prohibitively slow on real-world scores with thousands of notes. Alignment is necessitated by the lack of one-to-one correspondence labels.}.

MUSTER especially focuses on high-level accuracy and rhythmic structure, with sub-metrics for note-level edit-distance (\( \mathcal{E}_{\text{p}} \), \(\mathcal{E}_{\text{miss}} \), \( \mathcal{E}_{\text{extra}} \)), rhythm correction (\( \mathcal{E}_{\text{onset}} \)), defined by the amount of scale and shift operations required to correctly align every note's onset with the ground truth sequence, and \( \mathcal{E}_{\text{offset}} \), which measures the accuracy of the predicted note's musical durations.  
While edit-distance metrics primarily reflect the melodic correctness of a score, $\mathcal{E}_{\text{onset}}$ and $\mathcal{E}_{\text{offset}}$ serve as good indicators of rhythmic understanding and visual clarity of the resulting notation.

ScoreSimilarity also tracks edit-distances (\( \mathcal{E}_{\text{miss}} \), \( \mathcal{E}_{\text{extra}} \)) but additionally allows the evaluation of notational details such as stem direction ($\mathcal{E}_{\text{stem}}$), pitch spelling ($\mathcal{E}_{\text{spell.}}$), or hand/staff assignment ($\mathcal{E}_{\text{staff}}$). 
We extend ScoreSimilarity to ornaments by adding F1-scores for grace, staccato, and trill.
To harmonize the scores reported by both metrics, we opt to report normalized error scores and F1-scores instead of absolute error counts as originally proposed in \cite{cogliati2017metric}. 
\vspace{-0.6em}
\setlength{\tabcolsep}{2pt}

\begin{table}[!h]
\centering
\begin{tabular}{lrrrrrrr}
\toprule
& \multicolumn{6}{c}{\textbf{ScoreSimilarity} \cite{cogliati2017metric,suzuki2021score}}\\
\cmidrule(lr){2-7}
& \multicolumn{1}{c}{$\boldsymbol{\mathcal{E}_\text{staff}}$} 
& \multicolumn{1}{c}{$\boldsymbol{\mathcal{E}_\text{stem}}$} 
& \multicolumn{1}{c}{$\boldsymbol{\mathcal{E}_\text{spell.}}$} 
& \multicolumn{1}{c}{${\textbf{F1}_\text{grace}}$} 
& \multicolumn{1}{c}{${\textbf{F1}_\text{staccato}}$}
& \multicolumn{1}{c}{${\textbf{F1}_\text{trill}}$}
\\
\midrule
\makecell{SOTA \\ \cite{liu2022performance,MuseScore2023,FinaleV27}} & 6.86 & 26.74 & 9.69 & - & - & - &\\
\textbf{Ours} & \textbf{6.62} & \textbf{25.03} & \textbf{8.69} &  \textbf{27.80} & \textbf{18.19} & \textbf{54.64} \\
\bottomrule
\end{tabular}
\vspace{-0.6em}
\caption{Predicting score ornaments and visual details. 
}
\label{tab:ornaments-results}
\end{table}
\setlength{\tabcolsep}{6pt}

\vspace{-1.7em}
\setlength{\tabcolsep}{2.3pt}

\begin{table}[!h]
\centering
\begin{tabular}{lrrrrrr}
\toprule
& & \multicolumn{5}{c}{\textbf{ScoreSimilarity} \cite{cogliati2017metric,suzuki2021score}}\\
\cmidrule(lr){3-7}
\textbf{Method} & \( \boldsymbol{L} \) & \( \boldsymbol{\mathcal{E}_\text{miss}} \) & \( \boldsymbol{\mathcal{E}_\text{extra}} \) & \( \boldsymbol{\mathcal{E}_\text{duration}} \) & \( \boldsymbol{\mathcal{E}_\text{staff}} \) & \( \boldsymbol{\mathcal{E}_\text{stem}} \) \\ 
\midrule
Suzuki \cite{suzuki2021score} & 12954 & 12.53 & 4.21 & \textbf{0.53} & 0.03 & 5.40 \\
Octuple \cite{zeng2021musicbert} & 3697 & 3.56 & 4.74 & 17.51 & 16.34 & 30.94 \\
\textbf{Ours} & \textbf{3697} & \textbf{2.64} & \textbf{0.40} & 3.72 & \textbf{0.01} & \textbf{1.54} \\
\midrule
MIDI score\tablefootnote{The MIDI files were created from MusicXML with MuseScore 4.0.} & - & 3.04 & 4.64 & 13.63 & 3.41 & - \\
MusicXML & - & 0.00 & 0.00 & 0.00 & 0.00 & 0.00 \\ 
\bottomrule
\end{tabular}
\vspace{-0.6em}
\caption{Comparison of score representation schemes by sequence lengths and representation error rates.}
\label{tab:score-similarity-tokens}
\vspace{-1.0em}
\end{table}
\setlength{\tabcolsep}{6pt}

\subsection{Comparative experiments}
\hspace{1.3em}\textbf{PM2S.}\;
In \tabref{tab:unified-comparative}, we compare our model to the best publicly available PM2S systems. 
Our baselines include the popular commercial programs MuseScore \cite{MuseScore2023} and Finale \cite{FinaleV27}, the strongest HMM-based approach \cite{shibata2021non}, and the highest-performance deep learning model \cite{liu2022performance}, which relies on neural beat tracking.
Where necessary for evaluation, MuseScore 4 is employed to convert quantized score MIDI predictions to MusicXML. 
We also compare with an improved version of the \href{https://github.com/cheriell/PM2S}{reference implementation} of \cite{liu2022performance}, which removes the time-signature\footnote{We found that assuming a fixed $\frac{4}{4}$ time signature improves results.} and note-value prediction modules. 
However, as noted in Section \ref{subsub:datasets}, beat-tracking still struggles on real-world P-MIDI, lagging behind \cite{shibata2021non} and other options in rhythm quantization.

In contrast, our method predicts notation with significantly more accurate rhythm ($\mathcal{E}_{\text{onset}/\text{offset}}$), note values ($\mathcal{E}_{\text{offset}/\text{duration}}$), and fewer extraneous notes ($\mathcal{E}_{\text{extra}}$). 
In practice, this is reflected in better alignment of notes with barlines and more concise notation than alternative approaches.
While all baselines pass the input pitch sequence directly to the output, our setup requires the model to rebuild the full sequence from scratch, leading to more missed notes ($\mathcal{E}_{\text{p}/\text{miss}}$).
Decoupling the output pitch sequence from the input is key to our method, enabling training without one-to-one correspondences and predicting many-to-one relationships like trills.
In fact, many `misses' occur because our approach notated a trill where the ground truth score contains multiple alternating notes, with minimal impact on the resulting score's quality from a human perspective.

For sample scores and visual comparisons with baseline approaches, we refer to the supplementary material. 

\textbf{Notation details.}\;
To our knowledge, our method is the first PM2S system to predict note-level attributes beyond timing, pitch, and staff assignment.
The model also estimates staccato, grace notes, and trill marks, which are crucial for human performers. 
Given the data imbalance -- for instance, trills account for only 0.15\% of notes -- achieving high F1 scores is extremely challenging.
\tabref{tab:ornaments-results} shows that our approach predicts more accurate stem directions, pitch-spelling, and staff assignments, while exhibiting relatively good performance on grace and trill notes. 

\textbf{Tokenization scheme.}\label{subsub:musicxml-tokenization}
\setlength{\tabcolsep}{5.9pt}
\begin{table*}[!h]
\centering
{\renewcommand{\arraystretch}{0.985}
\begin{tabular}{@{\makebox[1.3em][r]{\textcolor{pink}{\rownumber}\space\space}}lrrrrrrrrrrr}
\toprule
 & \multicolumn{6}{c}{\textbf{MUSTER} \cite{shibata2021non}} & \multicolumn{5}{c}{\textbf{ScoreSimilarity} \cite{cogliati2017metric,suzuki2021score}} \\
\cmidrule(lr){2-7} \cmidrule(lr){8-12}
\gdef\rownumber{\stepcounter{magicrownumbers}\arabic{magicrownumbers}}
\textbf{Modification} & \( \boldsymbol{\mathcal{E}_\text{p}} \) & \( \boldsymbol{\mathcal{E}_\text{miss}} \) & \( \boldsymbol{\mathcal{E}_\text{extra}} \) & \( \boldsymbol{\mathcal{E}_\text{onset}} \) & \( \boldsymbol{\mathcal{E}_\text{offset}} \) & \( \boldsymbol{\mathcal{E}_\text{avg}} \) & \( \boldsymbol{\mathcal{E}_\text{miss}} \) & \( \boldsymbol{\mathcal{E}_\text{extra}} \) & \( \boldsymbol{\mathcal{E}_\text{duration}} \) & \( \boldsymbol{\mathcal{E}_\text{staff}} \) & \( \boldsymbol{\mathcal{E}_\text{stem}} \) \\
\midrule
no data augmentation\rowlabel{row:data-augmentation} & 10.96 & 38.05 & 38.23 & 37.89 & 51.34 & 35.30 & 49.99 & 47.98 & \textcolor{gray}{32.43} & \textcolor{gray}{7.30} & \textcolor{gray}{13.10} \\
BiLSTM backbone \rowlabel{row:lstm}& 4.01 & 22.28 & 16.02 & 38.30 & 60.30 & 28.12 &26.75 & 16.29 & \textcolor{gray}{55.46} &\textcolor{gray}{7.55}& \textcolor{gray}{21.08}\\
BiGRU backbone \rowlabel{row:gru} & 3.85 & 19.49 & 13.56 & 30.98 & 49.59 & 23.55 & 23.84 & 14.39 & \textcolor{gray}{43.69} & \textcolor{gray}{7.50} & \textcolor{gray}{22.33} \\
no beat-alignment\rowlabel{row:beat-alignment} & 3.57 & 25.53 & 10.91 &19.25 & 29.60 &17.77 & 33.40 & 10.91 & \textcolor{gray}{41.06} & \textcolor{gray}{6.03} & \textcolor{gray}{20.20} \\
no transpose\rowlabel{row:transpose} & 4.86 & 12.04 & 9.89 & 19.36 & 29.66 & 15.16 & 17.99& 12.76 & 50.54 & 7.92 & \textbf{24.93} \\
no 24-div quantization\rowlabel{row:non-musical-quantization} & \textbf{3.02} & 10.96 & 8.99 & 19.71 & 31.73 &14.88 & 15.43 & 10.50 & 74.92 & 7.54 & 26.84 \\
ALiBi pos. enc.\rowlabel{row:alibi}  & 3.12 & 11.22 & 8.07 &17.85 &27.66 & 13.58 & 16.24 & 10.11 & 55.67 & 7.35 & 27.92 \\
no surrogate pitch\rowlabel{row:surrogate-pitch} &4.11 & 9.83 & 8.20 & 16.89 & 26.66 & 13.14 & 15.21 & 10.79 & \textbf{49.41} & 6.81 & 25.84 \\
no onset jitter\rowlabel{row:onset-jitter} & 3.20 & 8.66 & 7.23 & 17.04 & 27.75 & 12.78 & 13.42 & 9.38 & 51.33 & 8.59 & 25.93\\
no tempo augmentation\rowlabel{row:tempo-augmentation} & 3.18 & 8.74 & 7.21 & 17.25 & 27.45 & 12.76 & 13.60 & 9.50 & 54.20 & 7.78 & 26.95 \\
no conditioning token\rowlabel{row:conditioning-token} & 3.09 & 9.10 & 7.10 & 16.82 & 27.01 & 12.62 &13.98 &9.18 & 52.98 & 8.33 & 26.57 \\
no duration jitter\rowlabel{row:duration-jitter} & 3.39 & 8.53 & 7.32 & 16.54 & 26.92 & 12.54 & 13.56 & 9.59 & 50.68 & 8.19 & 27.69\\
sinusoidal pos. enc.\rowlabel{row:sinusoidal-pos-enc} & 3.50 & 8.29 & 6.83 & 16.71 & 27.35 & 12.49 & 13.41 & 9.36 & 51.68 & 7.35 & 25.81 \\
\midrule
\textbf{Ours} \rowlabel{row:ours}& 3.11 & \textbf{7.56} & \textbf{6.44} & \textbf{15.55} & \textbf{23.84} & \textbf{11.30} & \textbf{12.69} & \textbf{9.06} & 51.86 & \textbf{6.62} & 25.03 \\
\bottomrule
\end{tabular}
}
\vspace{-1em}
\caption{Ablation study for key design decisions. \textcolor{gray}{Grayed out} values do not reflect the true model performance as a large fraction of notes are misaligned during metric computations, leading to incorrect results. Rows are organized by $\boldsymbol{\mathcal{E}_\text{avg}}$.
}
\label{tab:unified-ablation}
\vspace{-1.4em}
\end{table*}
\setlength{\tabcolsep}{6pt}
\hspace{-0.8em}We evaluate our MusicXML tokenization against prior methods and score-derived MIDI files by converting ground-truth scores to a new format and then comparing the reconstructions to the originals.

\tabref{tab:score-similarity-tokens} shows that our approach yields $3.5\times$ shorter sequence lengths than prior MusicXML tokenizations while maintaining more detail than alternatives.
Furthermore, it highlights MIDI's shortcomings as a notation format; both ground-truth MIDI scores and MIDI-based tokenizations \cite{zeng2021musicbert} exhibit lower fidelity than MusicXML-derived tokenizations and particularly high error rates for details like stem directions, which are not supported by MIDI. 

\vspace{-1.1em}
\subsection{Ablation study}\label{sec:ablation}
\vspace{-0.3em}
Our ablation study in \tabref{tab:unified-ablation} shows the impact of key design choices. 

\textbf{Backbone architecture.} The transformer architecture is much stronger than classic recurrent networks like bidirectional LSTMs \cite{hochreiter1997long} and GRUs \cite{chung2014empirical} when trained on the same data (rows \textcolor{pink}{\ref{row:gru}} \& \textcolor{pink}{\ref{row:lstm}}).
We also evaluate the effectiveness of the conditioning token (row \textcolor{pink}{\ref{row:conditioning-token}}) and demonstrate the impact of feeding surrogate pitch information to the encoder for unpaired data (row \textcolor{pink}{\ref{row:surrogate-pitch}}).

\textbf{Positional encoding.} We compare rotary embeddings with standard sinusoidal embeddings \cite{vaswani2017attention} (row \textcolor{pink}{\ref{row:sinusoidal-pos-enc}}) and ALiBi \cite{press2021train} (row \textcolor{pink}{\ref{row:alibi}}) and find that sinusoidal embeddings perform slightly weaker than rotary encoding while ALiBi yields significantly worse results.

\textbf{Tokenization.} We demonstrate the impact of our tokenization scheme's quantization by changing the encoding scheme to quantize durations by 32nd-divisions instead of 24th (row \textcolor{pink}{\ref{row:non-musical-quantization}}).
This has a particularly strong impact on metrics for rhythm and note values ($\mathcal{E}_\text{onset}$, $\mathcal{E}_\text{offset}$, $\mathcal{E}_\text{duration}$).

\textbf{Data augmentation \& alignment.} Data augmentation is crucial to the effectiveness of our approach (row \textcolor{pink}{\ref{row:data-augmentation}}).
Rows \textcolor{pink}{\ref{row:transpose}}, \textcolor{pink}{\ref{row:onset-jitter}}, \textcolor{pink}{\ref{row:tempo-augmentation}}, and \textcolor{pink}{\ref{row:duration-jitter}} show that using all 4 augmentation types combined yields the best results. 
We also demonstrate that using our greedy beat-level note alignment algorithm significantly improves performance compared to unoptimized input and output sequence alignment (row \textcolor{pink}{\ref{row:beat-alignment}}).

\subsection{Scaling}
\vspace{-0.3em}
To assess model performance as datasets grow, we conduct experiments with varying amounts of paired and unpaired data (see \tabref{tab:scaling-score-similarity}). 
We observe a clear trend where increasing the amount of paired and unpaired data improves final performance across most metrics.
The benefits of adding 10,000 unpaired scores are similar to expanding from 100 paired to 822 paired training pieces.
While training on unpaired data is less sample-efficient, the labeling cost-savings may make it worthwhile nonetheless. 
The results suggest that our method is able to leverage additional data well and that training on larger (paired \& unpaired) datasets could lead to significant further improvements.
\vspace{-.6em}
\setlength{\tabcolsep}{3pt}

\begin{table}[h]
\centering
\begin{tabular}{llrrrrr}
\toprule
\multicolumn{2}{c}{\textbf{Number of pieces}} & \multicolumn{5}{c}{\textbf{ScoreSimilarity} \cite{cogliati2017metric}} \\
\cmidrule(lr){1-2} \cmidrule(lr){3-7}
\textbf{Paired} & \textbf{Unpaired} & \multicolumn{1}{c}{$\boldsymbol{\mathcal{E}_\text{miss}}$} & \multicolumn{1}{c}{$\boldsymbol{\mathcal{E}_\text{extra}}$} & \multicolumn{1}{c}{$\boldsymbol{\mathcal{E}_\text{duration}}$} & \multicolumn{1}{c}{$\boldsymbol{\mathcal{E}_\text{staff}}$} & \multicolumn{1}{c}{$\boldsymbol{\mathcal{E}_\text{stem}}$} \\
\midrule
100 & - & 22.04 & 17.78 & 54.22 & 8.77 & 28.43 \\
822 & - & 14.44 & 10.11 & 50.77 & 7.78 & 27.56 \\
\midrule
100 & 10,000 & 15.40 & 12.03 & 55.98 & 7.53 & 26.93 \\
822 & 10,000 & 13.44 & 9.49 & \textbf{49.18} & 8.22 & 27.85 \\
822 & 58,686        & \textbf{12.69} & \textbf{9.06} & 51.86 & \textbf{6.62} & \textbf{25.03} \\
\bottomrule
\end{tabular}
\vspace{-0.4em}
\caption{The effects of training dataset size.}
\label{tab:scaling-score-similarity}
\vspace{-1.4em}
\end{table}
\setlength{\tabcolsep}{6pt}
\vspace{-0.6em}
\section{Conclusion}
We presented a flexible, robust, and conceptually simple approach to convert P-MIDI into musical notation and showed that a standard sequence-to-sequence transformer model can outperform existing methods that benefit from extensive domain-specific optimizations.
We also introduced a compact tokenization method for symbolic music data that is extensible to further notational elements in the future. 
Furthermore, we demonstrate that leveraging large unpaired training datasets can improve model performance and enhance the fidelity of predicted scores. 
Future efforts could add features like explicit key and time signature prediction, tempo marks, and expand to other musical genres, instruments, and notation systems.

\section{Ethics Statement}
The proposed system is primarily trained on classical piano music by European composers engraved in standard Western musical notation. 
While preliminary experiments show that the method generalizes out-of-genre to modern piano pop music, our approach nonetheless excludes a large body of musical work notated in different formats. 
Future work should aim to address this imbalance and create systems that can be useful to an even wider audience.

\section{Acknowledgements}
We would like to thank Andrew McLeod for his assistance and feedback on a draft of this work via the New-to-ISMIR paper mentoring program. This work was partially supported by the ERC Starting Grant SpatialSem (101076253).
\bibliography{ISMIRtemplate}
\clearpage
\end{document}